\documentclass[twocolumn,secnumarabic, amssymb, superscriptaddress, nobibnotes, aps, prl]{revtex4-2}
\usepackage{amsfonts}
\usepackage{amsmath}
\usepackage{amssymb}
\usepackage{appendix}
\usepackage{graphicx}%
\providecommand{\U}[1]{\protect\rule{.1in}{.1in}}
\providecommand{\U}[1]{\protect\rule{.1in}{.1in}}

\begin{document}
\title{Complete transparency with three active-passive-coupled optical resonators}
\author{Xiao-Bo Yan}
\affiliation{School of Physics and Electronic Engineering, Northeast Petroleum University, Daqing, 163318, China}
\author{Liu Yang}
\affiliation{College of Intelligent Systems Science and Engineering, Harbin Engineering University, Harbin, 150001, China}
\author{Bing He}
\email{bing.he@umayor.cl}
\affiliation{Center for Quantum Optics and Quantum Information, Universidad Mayor, Camino La Pir\'amide 5750, Huechuraba, Chile}

\begin{abstract}
The phenomena of induced transparency, with the typical examples of electromagnetically induced transparency (EIT) in atomic media and those based on coupled optical resonators, have attracted tremendous interest since their discoveries. Owing to the limitations of the involved physical elements, however, near-100\% transmissions were reported under highly demanding experimental conditions. With a structure of three linearly coupled optical resonators, an active one carrying optical gain and two passive ones simply with dissipation, we demonstrate that a transmitted light field can become completely transparent through the structure, which displays all properties similar to those of EIT. It is due to a destructive interference mechanism that totally eliminates the intracavity field in the dissipative resonator directly coupled to the transmitted field of any feasible power, when the coupling strength of two other resonators is tuned across a point determined by their associated gain and loss rates. This mechanism works for all possible coupling strengths of the dark resonator with the input field and its neighboring resonator, as well as for any available quality factor from its fabrication. The transparency window size and output field intensity can be freely adjusted by tuning two inter-cavity couplings of wide ranges, without modifying the built-in optical gain which can be just slightly stronger than the dissipation of the active resonator.
	\vspace{0.9cm} 
\end{abstract}
\maketitle

Since its first observation with a three-level atomic medium \cite{Harris1990prl,Boller1991prl}, the phenomenon of EIT has been extensively studied together with the accompanying slow-light effect \cite{Fleischhauer2005RMP,Ham1997,Hau1999Nat,Al2004,amplified,novikova-2012, EIT-review}. Due to non-zero decay rates of the involved atomic levels and additional absorption pathways in the presence of hyperfine or Zeeman sublevels, achieving a complete transparency under EIT condition is difficult. So far, near-$100$\% transmissions were reported with very few setups, such as optically dense rare-earth doped crystal \cite{Ham1997}, atomic vapor excited with circularly polarized laser of multimode \cite{Al2004}, and single artificial atoms of superconductor \cite{artificial-atom}. On the other hand, there exist the EIT-like phenomena in coupled optical resonators known as coupled-resonator-induced transparency (CRIT) \cite{cc-1,cc-2,cc-3,cc-5, microsphere,cc-6,cc-7,cc-8,cc-9,cc-10}, and they have developed into a prominent research direction \cite{report,review,trans-review}. Two linearly coupled dissipative resonators display all properties similar to those of three-level atomic EIT. The addition of nonlinear elements into optical resonator also leads to various types of induced transparency \cite{trans-review,q-001,q-0,dipole,q-01,q-1,q-2}. Especially, cavity optomechanical systems were found to demonstrate optomechanically induced transparency (OMIT) \cite{Huang2010_041803, Weis2010, Safavi-Naeini2011}, which has stimulated tremendous interests in its research \cite{Huang2011,Chang2011njp,Xiong2012,Tarhan2013,Karuza2013,Kronwald2013,Jing2015,Lu2018,Yan2020pra,Yan2021PhysicaE}. 
To these cavity systems, it is equally challenging to realize a complete transparency, due to the unavoidable dissipation of the involved elements such as optical resonator or mechanical resonator with a finite quality factor. 

Here, we present an experimentally feasible approach to realizing the complete transparency with three linearly coupled resonators [see Fig. 1(a)]. The structure consists of one active resonator with optical gain and two other dissipative ones.
A special cavity dark mode with totally vanishing intracavity field in the dissipative resonator directly coupled to the input emerges under certain conditions, and it realizes the 100\% transmission at the center of transparency window. Unlike various schemes \cite{amplified,cc-10,cc-12} where optical gain is used to compensate for optical loss, the complete intracavity field cancellation is realizable with an arbitrary net gain in the neighboring resonator, no matter how large the field loss in the dark cavity can be. Contrary to intuition, it is not necessary to have an correspondingly enhanced gain to balance any amount of increased field damping for the dark cavity, as long as the inter-cavity coupling between two other resonators can be adjusted over a suitable range. Different from atomic EIT and many other induced-transparency scenarios such as CRIT and OMIT, this phenomenon is based on a previously unknown interference mechanism that is irrespective of the transmitted field intensity and the coupling strengths of the others to the dark resonator. 

For the structure shown in Fig. 1(a), the symbols $\hat{a}_{i}$ and $\omega_{i}$ stand for the intracavity mode and the resonance frequency of cavity $i$ ($i=1,2,3$), respectively. 
The transmitted field at the frequency $\omega_{p}$ and with the amplitude $\varepsilon_{p}=\sqrt{P/(\hbar\omega_p)}$ ($P$ is its power) couples to cavity $1$. The active resonator, cavity $2$, has the net 
gain rate $\kappa_{2}=\kappa_{2,0}/\left(1+\frac{<\hat{a}^\dagger_{2}\hat{a}_{2}>}{I_S}\right)-\gamma_2>0$ ($\kappa_{2,0}$ is the initial gain rate, $I_S$ the gain saturation intensity, and $\gamma_2$ the cavity's damping rate) and is coupled to two other dissipative ones with their damping rates $\kappa_1$ and $\kappa_3$, respectively, at the inter-cavity coupling strengths $J_{1}$ and $J_{2}$, which are determined by the distances between the resonators. Hence, similar to those of parity-time (PT) symmetric optical structures \cite{pt-2,pt-3,pt-4,pt-5,pt-6,pt-7}, the effective Hamiltonian of the setup reads as ($\hbar=1$)
\begin{eqnarray}
	\hat{H}_{eff}&=&\omega_{1}\hat{a}_{1}^{\dagger}\hat{a}_{1}+\omega_{2}\hat{a}_{2}^{\dagger}\hat{a}_{2}+\omega_{3}\hat{a}_{3}^{\dagger}\hat{a}_{3}-J_{1}(\hat{a}_{1}^{\dagger}\hat{a}_{2}+\hat{a}_{2}^{\dagger}\hat{a}_{1})\notag\\&-&J_{2}(\hat{a}_{2}^{\dagger}\hat{a}_{3}+\hat{a}_{3}^{\dagger}\hat{a}_{2})+i\sqrt{2\kappa_{ex}}\varepsilon_{p}(\hat{a}_{1}^{\dagger}e^{-i\omega_{p} t}-\hat{a}_{1}e^{i\omega_{p}t})\notag\\&-&i\kappa_1\hat{a}_{1}^{\dagger}\hat{a}_{1}+i\kappa_{2}\hat{a}_{2}^{\dagger}\hat{a}_{2}-i\kappa_3\hat{a}_{3}^{\dagger}\hat{a}_{3},
	\quad
	\label{Hamiltonian}   
\end{eqnarray}%
where the damping rate $\kappa_1=\kappa_{ex}+\kappa_{in}$ of cavity $1$ comes from both external coupling and intrinsic loss. 

\begin{figure}[t]
	\includegraphics[width=0.45\textwidth]{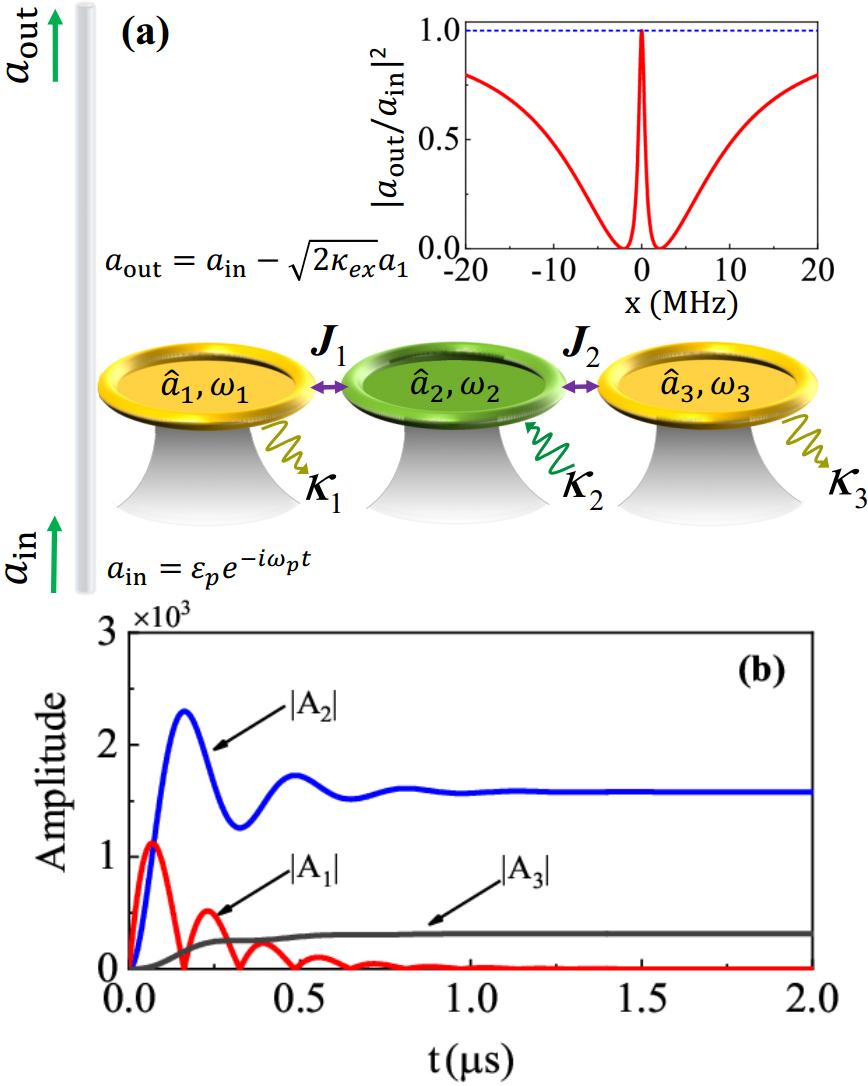}
	\caption{(a) The schematic structure of three linearly coupled microresonators. Resonator $2$ carries a gain medium without showing its associated pump. The actual position of the optical fiber is within the plane of the microresonators. There will be a totally vanishing intracavity field $\hat{a}_1=0$, once resonator $2$, $3$ and a single-frequency input satisfy the conditions in Eq. (\ref{g-condiitons}). Then, the transmitted fields will have the $100$\% transmission at a transparency window center, as exemplified in the inset obtained with $\kappa_{1}=2\kappa_{ex}=10\,\mathrm{MHz}$, $\kappa_{2}=0.2\,\mathrm{MHz}$, $\kappa_{3}=5\,\mathrm{MHz}$, $J_1=2\,\mathrm{MHz}$, and $J_2=1\,\mathrm{MHz}$, for three microresonators of the same resonance frequency. (b) An example of the evolving intracavity fields under gain saturation effect. The parameters are $\varepsilon_p=10^4\,\sqrt{\mathrm{MHz}}$ (equivalent to $P=12.8\, \mathrm{\mu W}$ at $\lambda=1550$ nm), $\kappa_{1}=2\kappa_{ex}=10\, \mathrm{MHz}$, $\kappa_{2,0}=0.2\, \mathrm{MHz}$, $I_S=10^8$, $\kappa_3= 5\, \mathrm{MHz}$, $J_1=20\, \mathrm{MHz}$, and $J_2=1\, \mathrm{MHz}$. The saturated gain rate after the stabilization is $\kappa_{2}=0.195\, \mathrm{MHz}$, and the cavity damping rates are chosen to be on the levels of those in some recent experiments \cite{mc1,mc2,mc3}.}%
	\label{Fig1}%
\end{figure}

To capture the main physical picture, we will consider a constant net gain rate $\kappa_{2}$ while the system is driven by a transmitted field. In addition to the previous studies (see e.g., Refs. \cite{He2018prl,sature1,sature2,sature-nonreciprocal,sature3}), the effect of gain saturation with a real-time $\kappa_{2}(t)$ during the system's dynamical evolution [like the process in Fig. 1(b)], together with the associate gain noise effect \cite{gain-noise,sature4} which can be neglected in experiments \cite{ex-pt1}, is discussed in Appendix A and D.
Then the effective Hamiltonian in Eq. (1) leads to the following dynamical equation of the  
mean modes $a_{i}(t)=\langle\hat{a}_{i}(t)\rangle$ ($i=1,2,3$):
\begin{equation}
	\dot{\vec{\textit{\textbf{v}}}}=\textbf{M}\vec{\textit{\textbf{v}}}+\vec{\textit{\textbf{v}}}_{in}%
\end{equation}
with $\vec{\textit{\textbf{v}}}=(a_{1},a_{2},a_{3})^{\mathrm{T}}$, $\vec{\textit{\textbf{v}}}_{in}=(\sqrt{2\kappa_{ex}}\varepsilon_{p}e^{-i\omega_{p}t},0,0)^{\mathrm{T}}$, and
\begin{equation}
	\textbf{M}=\left(
	\begin{array}
		[c]{ccc}%
		-\kappa_{1}-i\omega_{1} & iJ_{1} & 0\\
		iJ_{1} & \kappa_{2}-i\omega_{2} & iJ_{2}\\
		0 & iJ_{2} & -\kappa_{3}-i\omega_{3}
	\end{array}
	\right).
	\label{matrix}
\end{equation}

In a stable regime (the real parts of all eigenvalues of $\textbf{M}$ are not positive), the solution to Eq. (2) takes the form $a_{k}=A_{k}e^{-i\omega_{p}t}$ ($k=1,2,3$). Substituting this form into Eq. (2), we obtain the exact solution:
\begin{eqnarray}
	A_{1}&=&\dfrac{\sqrt{2\kappa_{ex}}\varepsilon_{p}[J_{2}^{2}-(\kappa_{2}+i\Delta_{2})(\kappa_{3}-i\Delta_{3})]}{\eta+[J_{1}^{2}-(\kappa_{1}-i\Delta_{1})(\kappa_{2}+i\Delta_{2})](\kappa_{3}-i\Delta_{3})}\notag\\
	A_{2}&=&\dfrac{\sqrt{2\kappa_{ex}}\varepsilon_{p}J_{1}(\Delta_{3}+i\kappa_{3})}{\eta+[J_{1}^{2}-(\kappa_{1}-i\Delta_{1})(\kappa_{2}+i\Delta_{2})](\kappa_{3}-i\Delta_{3})}\notag\\
	A_{3}&=&\dfrac{-\sqrt{2\kappa_{ex}}\varepsilon_{p}J_{1}J_{2}}{\eta+[J_{1}^{2}-(\kappa_{1}-i\Delta_{1})(\kappa_{2}+i\Delta_{2})](\kappa_{3}-i\Delta_{3})},\quad 
\end{eqnarray} 
with $\Delta_{i}=\omega_{p}-\omega_{i}$ ($i=1,2,3$) and $\eta=J_{2}^{2}(\kappa_{1}-i\Delta_{1})$. $A_{2}$ and $A_{3}$ cannot be zero because $J_{1}$, $J_{2}$ and $\kappa_{3}$ are non-zero, but $A_{1}$ can be zero under certain conditions. This is a unique feature of the structure in Fig. 1(a) and is crucial for achieving a complete transparency. To see the conditions for complete transparency more clearly, we rewrite $A_{1}$ in the form of continued fraction:
\begin{eqnarray}
	A_{1}=\dfrac{\sqrt{2\kappa_{ex}}\varepsilon_{p}}{\kappa_{1}-i\Delta_{1}-\dfrac{J_{1}^{2}}{\kappa_{2}+i\Delta_{2}-\dfrac{J_{2}^{2}}{\kappa_{3}-i\Delta_{3}}}}.
\end{eqnarray}%
Such continued fraction forms help to find the complete transparency conditions for all similar structures (see Appendix B). The properties of a transmitted light field is determined by the induced $A_1$, with the real and imaginary part of the susceptibility $\varepsilon_{T}=\sqrt{2\kappa_{ex}}A_{1}/\varepsilon_{p}$ (reformulated from a similar definition \cite{Huang2010_041803} by our form of dynamical equation) respectively describing the absorptive and dispersive behavior of a single-frequency input to the structure.  

Making the denominator of one subfraction in Eq. (5) vanish, i.e.,
\begin{eqnarray}
	\kappa_{2}+i\Delta_2-\dfrac{J_{2}^{2}}{\kappa_{3}-i\Delta_3}=0,
\end{eqnarray}
will lead to a complete transparency.
Under this condition, one has $a_1=0$, and the input-output relation in Fig. 1(a) indicates the exactly same output  $a_{out}$ as the input $a_{in}=\varepsilon_p e^{-i\omega_pt}$.
Since both real and imaginary parts of Eq. (6) should be zero, it follows that
\begin{eqnarray}
	J_{2}&=&\sqrt{\dfrac{\kappa_{2}\kappa_{3}[(\omega_{2}-\omega_{3})^{2}+(\kappa_{2}-\kappa_{3})^{2}]}{(\kappa_{2}-\kappa_{3})^{2}}}\notag\\
	\omega_{p}&=&\dfrac{\omega_{3}\kappa_{2}-\omega_{2}\kappa_{3}}{\kappa_{2}-\kappa_{3}}.
	\label{g-condiitons}
\end{eqnarray}%
Now we consider three microresonators with the same resonance frequency $\omega_{1}=\omega_{2}=\omega_{3}=\omega_{0}$ from their fabrication and the control of their temperatures \cite{ex-pt1}, so that Eq. (7) will be simplified to
\begin{eqnarray}
	J_{2}&=&\sqrt{\kappa_{2}\kappa_{3}}\notag\\
	\omega_{p}&=&\omega_{0}, 
	\label{conditions}
\end{eqnarray}%
and the susceptibility becomes
\begin{eqnarray}
	\varepsilon_{T}=\dfrac{2\kappa_{ex}}{\kappa_{1}-ix-\dfrac{J_{1}^{2}}{\kappa_{2}+ix-\dfrac{\kappa_{2}\kappa_{3}}{\kappa_{3}-ix}}}, 
	\label{susceptibility}
\end{eqnarray}%
where $x=\omega_{p}-\omega_{0}$. In comparison, the susceptibility of two coupled passive resonators \cite{cc-8,review,trans-review} takes the form $\varepsilon_{T}=2\kappa_{ex}/[\kappa_{1}-ix+J^{2}/(\gamma_{2}-ix)]$, with the corresponding definitions for the coupling and damping rates. The denominator $\gamma_{2}-ix$ of its subfraction cannot be zero, due to the second cavity's damping rate $\gamma_2\neq 0$ in reality. 

\begin{figure}[t]
	\centering
	\includegraphics[width=0.49\textwidth]{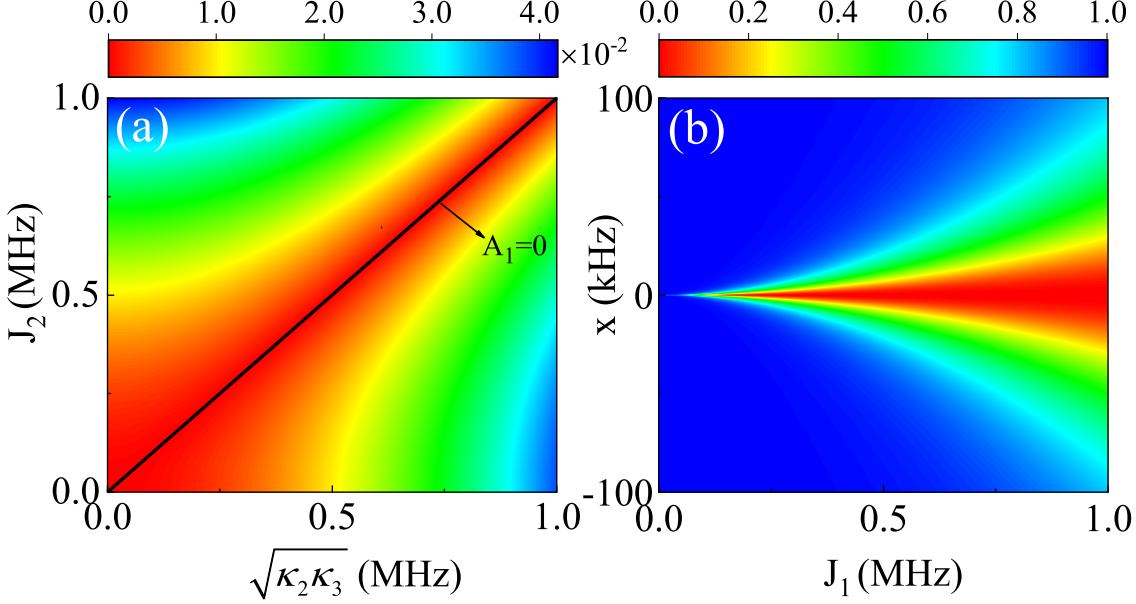}\quad 
	\caption{(a) The distribution of the absolute susceptibility, $|\varepsilon_T|=\sqrt{2\kappa_{ex}}|A_{1}|/\varepsilon_{p}$, at $x=0$, with respect to two parameters $\sqrt{\kappa_2\kappa_3}$ and $J_2$. Here, we fix the rest parameters at $\kappa_1=2\kappa_{ex}=20\, \mathrm{MHz}$, $\kappa_3=5\, \mathrm{MHz}$, and $J_1=10\, \mathrm{MHz}$. The calculation is based on Eq. (5) but with the same cavity resonance frequency. (b) The distribution of $\text{Re}(\varepsilon_T)$ with respect to $x$ and $J_1$. Here, the rest of the parameters are set to be $\kappa_1=2\kappa_{ex}=20\, \mathrm{MHz}$, $\kappa_2=0.05\, \mathrm{MHz}$, $\kappa_3=20\, \mathrm{MHz}$, and $J_2=1\, \mathrm{MHz}$. The calculation is based on Eq. (9).}%
	\label{Fig2}%
\end{figure}

The stabilized intracavity fields,
\begin{eqnarray}
	(A_1,A_2,A_3)=\sqrt{2\kappa_{ex}}\varepsilon_p\left(0,\frac{i}{J_1},-\sqrt{\frac{\kappa_2}{\kappa_3}}\frac{1}{J_1}\right)
	\label{state}
\end{eqnarray}
under the conditions of complete transparency in Eq. (\ref{conditions}), correspond to the zero eigenstate of the effective Hamiltonian in Eq. (\ref{Hamiltonian}) (see the proof in Appendix C). Similar to the definition for a network of linearly coupled bosonic modes \cite{ds0}, the mode $\hat{a}_1$ with $A_1=0$, which is decoupled from all possible excitations of $\hat{a}_2$ and $\hat{a}_3$, is regarded as a cavity dark mode. Different from those formed by atom-photon interactions \cite{a-dm01,a-dm02,a-dm03} and optomechanical interaction \cite{om-d1,om-d2}, this cavity dark mode is simply created by the linear couplings between the cavities, together with an optical gain in one of them. In this vacuum state $|0\rangle$ of cavity $1$, the driving action $\hat{H}_1=i\sqrt{2\kappa_{ex}}\varepsilon_{p}(\hat{a}_{1}^{\dagger}-\hat{a}_1)$ of the transmitted field and the coupling action $\hat{H}_2=-J_1(A_2\hat{a}_1^\dagger+A_2^{\ast}\hat{a}_1)$ from cavity $2$ could affect its photon number (see Appendix C). However, one has the exact cancellation 
\begin{eqnarray}
	\langle 1|\hat{H}_1+\hat{H}_2|0\rangle&=&\langle 1|i\sqrt{2\kappa_{ex}}\varepsilon_{p}\hat{a}_{1}^{\dagger}-J_1A_2\hat{a}_1^\dagger|0\rangle \nonumber\\
	&=& 0, 
\end{eqnarray}
with the stabilized $A_2=i\sqrt{2\kappa_{ex}}\varepsilon_p/J_1$ in Eq. (\ref{state}), indicating an absolutely forbidden transition to the excited states of cavity $1$. This perfect cancellation exists for all possible values of the directly involved system parameters ($J_1$ and $\sqrt{2\kappa_{ex}}\varepsilon_{p}$), in contrast to the destructive interference mechanisms in the previously known scenarios of induced transparency. Tuning the coupling strength $J_2$ (with the examples on microspheres \cite{microsphere}, ring resonators \cite{cc-10}, and microtoroid resonators \cite{cc-8,coupling}) so that the system goes close to the diagonal line in Fig. 2(a), one will obtain an approximate vacuum state of cavity $1$. Varying the other coupling $J_1$ will modify the distribution of this approximate dark mode with respect to the detuning $x$; see Fig. 2(b). The continued fraction form in Eq. (5) shows that no matter how small $J_{1}$ is, there exist the approximate dark modes in the neighborhood of the parameter submanifolds determined by Eq. (6). 

Once the mode $\hat{a}_1$ is tuned into a dark one, the fields in cavity $2$ and $3$ will exchange the photons with each other in a self-sustained way, having their photon numbers at the fixed ratio $\langle\hat{a}_2^\dagger\hat{a}_2\rangle/\langle\hat{a}_3^\dagger\hat{a}_3\rangle=\kappa_{3}/\kappa_2$ without any supply from the input field. The stabilized photon number $\langle\hat{a}_2^\dagger\hat{a}_2\rangle$ in the active resonator, 
which counter-intuitively becomes independent of its net gain rate $\kappa_2$ [see Eq. (\ref{state})], is unaffected by their coupling strength $J_{2,c}=\sqrt{\kappa_{2}\kappa_3}$ in this state. For any such system that can be tuned to this point $J_{2,c}$, the designed gain rate can be just slightly higher than the damping rate $\gamma_2$, 
to have a net gain rate $\kappa_2$ as small as possible with a correspondingly increased $\kappa_3$. 

The most important features of the structure are reflected in the absorptive curve of $\text{Re}(\varepsilon_T)$ and the dispersive curve of $\text{Im}(\varepsilon_T)$. In a situation of $\kappa_{2}<\kappa_{3}$, these curves after the coupling strength $J_2$ is tuned to $J_{2,c}$ are exemplified by those in Fig. 3(a1). The size of the transparency window, the full width at half maximum (FWHM) of the curve $\text{Re}(\varepsilon_T)$ around $x=0$, can be approximated as $J_1^2/\kappa_1$ when $J_1$ is small; cf. the approximate transparency window width $|\Omega|^2/\gamma$ ($\Omega$ is the pump Rabi frequency and $\gamma$ the level decay rate) of three-level atomic EIT. Correspondingly, the slope of the curve of $\text{Im}(\varepsilon_T)$ has the exact form,  
\begin{eqnarray}
	K=\frac{2\kappa_{ex}(\kappa_{2}-\kappa_{3})}{J_{1}^{2}\kappa_{3}}
	\label{slope}
\end{eqnarray}
at $x=0$, happening to be the opposite of the time delay $\tau=\partial\arg(a_{out}/a_{in})/\partial\omega_{p}$
of a transmitted field. In addition to the slow-light effect shown in Fig. 3(a2), the transparency window 
can be quickly shrunk by reducing the coupling $J_1$ and enhancing the $\kappa_{ex}$ to the overcoupling regime, to easily reach the sub-mHz range as the recently proposed nuclear spin induced transparency \cite{nuc-trans} can realize. 
For this system satisfying $\kappa_{2}<\kappa_{3}$ and $J_{2}=\sqrt{\kappa_{2}\kappa_{3}}$, it will keep its dynamical stability over the whole range $0<J_1<\infty$ for coupling cavity $1$ and $2$ (see Appendix D). The above results illustrated by the exactly solved curves of $\text{Re}(\varepsilon_T)$ and $\text{Im}(\varepsilon_T)$, which go through the zero point together under the complete transparency conditions, are irrelevant to the input field intensity, and it is a prominent difference from atomic EIT and OMIT where a transmitted field should be relatively weak.

\begin{figure}[t]
	\centering
	\includegraphics[width=0.49\textwidth]{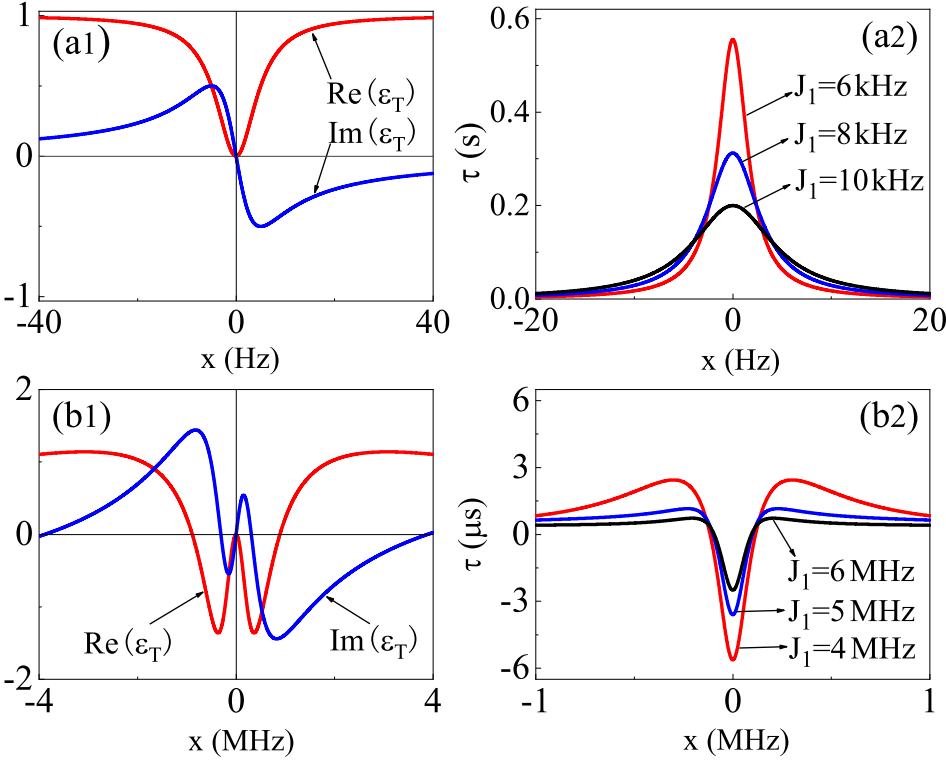}\quad 
	\caption{(a1)-(a2) The examples of $\text{Re}(\varepsilon_T)$ and $\text{Im}(\varepsilon_T)$, as well as the time delays of the transmitted fields, in the regime of $\kappa_{2}<\kappa_{3}$. We use $\kappa_1=2\kappa_{ex}=20\,\mathrm{MHz}$, $\kappa_2=0.01\,\mathrm{MHz}$, $\kappa_3=20\,\mathrm{MHz}$, $J_1=0.01\,\mathrm{MHz}$ in (a1), and $J_2=\sqrt{\kappa_{2}\kappa_{3}}$, so that the transparency window size is in the order of $\mathrm{Hz}$ and the time delays approach the order of second. (b1)-(b2) The other examples in the regime of $\kappa_{2}>\kappa_{3}$. The system parameters are set to be $\kappa_1=2\kappa_{ex}=10\,\mathrm{MHz}$, $\kappa_2=1\,\mathrm{MHz}$, $\kappa_3=0.1\,\mathrm{MHz}$, and $J_2=\sqrt{\kappa_{2}\kappa_{3}}$. In (b1) we have $J_1=4\,\mathrm{MHz}$. }%
	\label{Fig3}%
\end{figure}

There will be totally different response to the input, if the prepared gain rate $\kappa_2$ surpasses the damping rate $\kappa_{3}$ of a neighboring resonator but keeps the system to be dynamically stable (see Appendix D). Now the curve of $\text{Re}(\varepsilon_T)$ takes the form in Fig. 3(b1), showing a window centered at the minimum amplification. The corresponding curve of $\text{Im}(\varepsilon_T)$ has a slope $K>0$ at $x=0$, displaying  
a fast-light effect with the negatives time delays in Fig. 3(b2). 

\begin{figure}[h]
	\centering
	\includegraphics[width=0.5\textwidth]{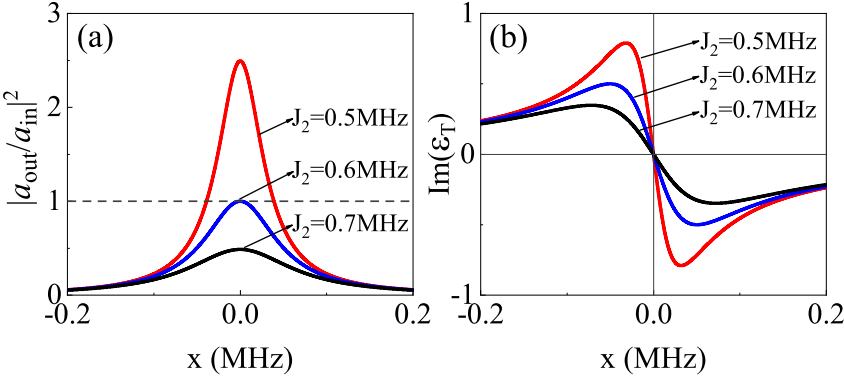}\quad 
	\caption{(a) The comparison between the output/input ratios due to three different couplings $J_2$, which are, respective smaller than (the red curve), equal to (the blue curve), and larger than (the black curve) the value $J_{2,c}$ of the $100\%$ transmission at the transparency window center. Represented by these three samples, the output can be continuously amplified or reduced by tuning the coupling $J_2$ across the point $J_{2,c}$, while the built-in optical gain is unchanged. (b) The corresponding dispersive curves. Here, the fixed system parameter are chosen as $\kappa_1=2\kappa_{ex}=20\,\mathrm{MHz}$, $\kappa_2=0.06\,\mathrm{MHz}$, $\kappa_3=6\,\mathrm{MHz}$, and $J_1=1 \,\mathrm{MHz}$.      }%
	\label{Fig4}%
\end{figure}

Instead of being a limit point that should be asymptotically approached, the completely vanishing field in cavity $1$ emerges at a midpoint of tuning the coupling $J_2$. As represented by the samples in Fig. 4(a), tuning the coupling strength from $J_2>J_{2,c}=\sqrt{\kappa_{2}\kappa_{3}}$ to $J_2<J_{2,c}$ will continuously switch the transparency window center from less than the $100\%$ transmission to more than the $100\%$ transmission, while the dispersion property demonstrating the slow-light effect is still preserved as seen from Fig. 4(b). As the distance between cavity $2$ and $3$ keeps increasing from where cavity $1$ is in the vacuum state of completely eliminated intracavity field, which is at $J_{2,c}=0.6$ MHz for the system in Fig. 4, 
more and more photons will enter cavity $1$ and realize an intracavity field phase for the mode $a_1$ to interfere with the input field constructively, so that a higher and higher ratio $|a_{out}/a_{in}|^2>1$ will be achieved until the system loses its dynamical stability at $J_2=0.2449$ MHz. Then it becomes something like two coupled resonators in a PT-symmetry broken regime due to the choice $J_1<\sqrt{\kappa_1\kappa_2}$. There is no restriction on the qualities of used optical resonators for seeing these phenomena of amplifying and reducing the output without modifying the built-in optical gain, since 
the system stability under an arbitrary net gain rate $\kappa_{2}$, which is detailed in Appendix D, and an available coupling strength range covering the point $J_{2,c}$ guarantee their occurrence. The structure in Fig. 1(a) can thus beat a common notion that the better the used optical resonators are, the better the induced transparency will be.

In summary, we have detailed an interference mechanism that genuinely reduces the susceptibility $\varepsilon_T$ of the structure in Fig. 1(a) to zero under the implementable conditions, so that a light field transmitted at the resonance frequency of those coupled resonators and with any feasible power will become totally invisible to the structure, no matter how strongly it is coupled to the first resonator. Besides scanning the transmitted field frequency, we only require the adjustments of two inter-cavity couplings to see its manifestation: setting one inter-cavity coupling at $J_{2,c}=\sqrt{\kappa_2\kappa_3}$, and independently choosing the other coupling $J_1$ for a suitable transparency window size. Rather than being used to cancel the transmission loss, the optical gain with its arbitrary net gain rate $\kappa_{2}$ creates the point of perfect interference in the first resonator, together with two other parameters $J_2$ and $\kappa_{3}$ that are also totally independent of the properties of cavity $1$, and can modify the dispersion of a transmitted field with the varied $\kappa_{2}$. The significance of the scenario lies in two aspects. First, to the research of induced transparency based on optical resonator \cite{review,trans-review}, it proves that the ultimate goals of complete transparency and free control on the output/input ratio can be realized without complex designs or particular nonlinear elements. Second, it will non-trivially extend the scope of non-Hermitian photonics based on optical gain. Optical gain in one of two coupled optical resonators provides a realization of PT symmetry \cite{pt-2,pt-3,pt-4,pt-5,pt-6,pt-7,ex-pt1,ex-pt2}, and the study of the associated exceptional points has been developed into a direction of intensive research \cite{ep2,ep3,ep4}. Coupling one more resonator will have an even more non-trivial scenario involving a special dark mode with its rich properties, and it presents a rare example of something unusual simply built on the common elements like active, passive resonators and their linear couplings.

{\it Acknowledgments}---This work was supported by the Program for Young Talents of Basic Research in Universities of Heilongjiang Province (YQJH2024037). L. Y. is supported by National Natural Science Foundation of China (62273115). B. H. is sponsored by ANID Fondecyt Regular (1221250) and thanks Dr. Xiaoshun Jiang for the discussion on the relevant experimental issues.

\vspace{-0.5cm}
\bigskip

\appendix

\begin{widetext}
	
	\renewcommand{\theequation}{S-\arabic{equation}}
	\setcounter{equation}{0}  
	
	\renewcommand{\thefigure}{S-\arabic{figure}}
	\setcounter{figure}{0}  
	
	\renewcommand{\thetable}{S-\arabic{table}}
	\setcounter{table}{0}  
	
	\section{Appendix}
	
	\subsection{A. Effects of Optical Gain Saturation and Amplification Noise}
	
	In non-Hermitian optical models, such as in parity-time (PT) symmetric photonics \cite{pt-3,pt-7}, the systems are often described by the effective non-Hermitian Hamiltonian with a constant optical gain rate. Here, we adopt another approach to the system dynamics from the full quantum view of stochastic Hamiltonian \cite{book}, which naturally includes the noise effects such as those from the optical amplification noise \cite{gain-noise,sature2,sature4}. Following the notation in Refs. \cite{sature2,sature4},
	we write the stochastic Hamiltonian as follows (the definitions for the parameters are the same as those in the main text 
	and $\hbar=1$):
	\begin{eqnarray}
		\hat{H}&=&\omega_{1}\hat{a}_{1}^{\dagger}\hat{a}_{1}+\omega_{2}\hat{a}_{2}^{\dagger}\hat{a}_{2}+
		\omega_{3}\hat{a}_{3}^{\dagger}\hat{a}_{3}-J_{1}(\hat{a}_{1}^{\dagger}\hat{a}_{2}+\hat{a}_{2}^{\dagger}\hat{a}_{1})-J_{2}(\hat{a}_{2}^{\dagger}\hat{a}_{3}+\hat{a}_{3}^{\dagger}\hat{a}_{2})+i\sqrt{2\kappa_{ex}}\varepsilon_{p}(\hat{a}_{1}^{\dagger}e^{-i\omega_{p} t}-\hat{a}_{1}e^{i\omega_{p}t})\notag\\&&+i\sqrt{2g_{2}(t)}(\hat{\xi}^\dagger\hat{a}_2^\dagger-\hat{\xi}\hat{a}_2)+i\sqrt{2\kappa_{in}}(\hat{\zeta}\hat{a}_1^\dagger-\hat{\zeta}^\dagger\hat{a}_1)+i\sqrt{2\gamma_{2}}(\hat{\zeta}\hat{a}_2^\dagger-\hat{\zeta}^\dagger\hat{a}_2)+i\sqrt{2\kappa_{3}}(\hat{\zeta}\hat{a}_3^\dagger-\hat{\zeta}^\dagger\hat{a}_3),
		\quad
		\label{full-Hamiltonian}   
	\end{eqnarray}%
	where the total losses in cavity $1$ give the damping rate $\kappa_{1}=\kappa_{in}+\kappa_{ex}$. By applying the Ito's rules for the gain noise operator $\hat{\xi}(t)$ and dissipation noise operator $\hat{\zeta}(t)$, we will obtain the associated dynamical equations of the intracavity field modes:
	\begin{eqnarray}
		\dot{\hat{a}}_{1}&=&-(\kappa_{1}+i\omega_{1})\hat{a}_1+iJ_1\hat{a}_2+\sqrt{2\kappa_{ex}}\varepsilon_{p}e^{-i\omega_pt}+\sqrt{2\kappa_{in}}\hat{\zeta},\notag\\
		\dot{\hat{a}}_{2}&=&iJ_1\hat{a}_1+(g_2(t)-\gamma_2-i\omega_2)\hat{a}_2+iJ_2\hat{a}_3+\sqrt{2g_{2}(t)}\hat{\xi}^\dagger
		+\sqrt{2\gamma_{2}}\hat{\zeta},\notag\\
		\dot{\hat{a}}_{3}&=&iJ_2\hat{a}_2-(\kappa_{3}+i\omega_{3})\hat{a}_3+\sqrt{2\kappa_{3}}\hat{\zeta},\quad 
		\label{equations}
	\end{eqnarray}
	where the noise operators satisfy the relations $\langle\hat{\xi}(t)\hat{\xi}^\dagger(t')\rangle=\delta(t-t')$ and $\langle\hat{\zeta}^\dagger(t)\hat{\zeta}(t')\rangle=0$. In some other literature the damping terms take the form $\kappa/2$ while the corresponding noise terms have the factor $\sqrt{\kappa}$. Particularly, we consider the gain saturation effect such that the net gain rate with time takes the form 
	\begin{eqnarray}
		\kappa_{2}(t)=\frac{\kappa_{2,0}}{1+\frac{<\hat{a}^\dagger_{2}\hat{a}_{2}(t)>}{I_S}}-\gamma_2=g_2(t)-\gamma_2,
		\label{gain}
	\end{eqnarray}
	where $\kappa_{2,0}$ is the initial gain rate when cavity $2$ is in the vacuum state, $I_S$ the gain saturation intensity determined by the medium and setup, and $\gamma_2$ the damping rate of cavity $2$.    
	
	\begin{figure}[h]
		\centering
		\includegraphics[width=1.0\textwidth]{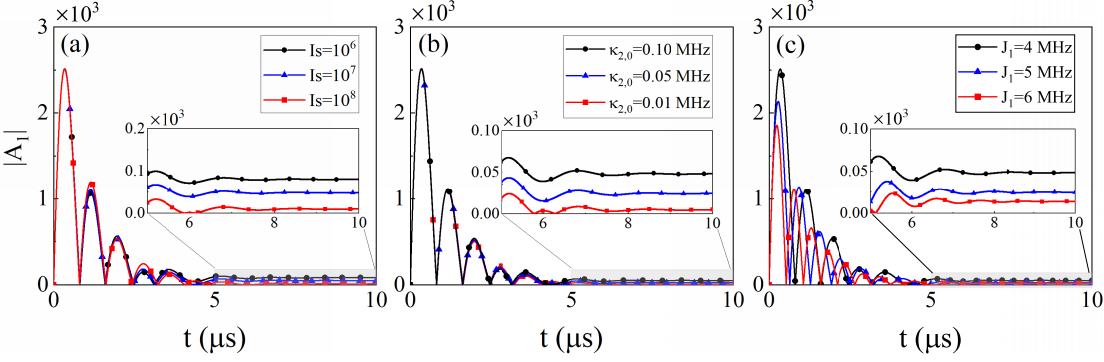}\quad 
		\caption{ (a) Examples of the evolution of the field amplitude $|A_1|$ for different saturation intensities $I_S$, while the other parameters of the setup are fixed.  Here, the initial gain rate is $\kappa_{2,0}=0.1\,\mathrm{MHz}$, and $J_{1}=4\,\mathrm{MHz}$. (b) Examples of the evolution of the field amplitude $|A_1|$ for different initial gain rates $\kappa_{2,0}$, while the other parameters of the setup are fixed. Here, the gain saturation intensity is fixed at $I_S=10^7$, and $J_{1}=4\,\mathrm{MHz}$. (c) Examples of the evolution of the field amplitude $|A_1|$ for different inter-cavity couplings $J_1$, while the other parameters of the setup are fixed. Here, we have the initial gain rate $\kappa_{2,0}=0.1\,\mathrm{MHz}$ and the gain saturation intensity $I_S=10^7$. In all these examples, the fixed system parameters are $\kappa_{1}=2\kappa_{ex}=2\,\mathrm{MHz}$ and $\kappa_{3}=1\,\mathrm{MHz}$. The pump field amplitude is fixed at $\varepsilon_p=10^4\,\sqrt{\mathrm{MHz}}$. }%
		\label{FigS1}%
	\end{figure}
	
	In some applications, gain saturation effect leads to optical isolation of coupled optical resonators \cite{He2018prl,sature-nonreciprocal, ex-pt1}, modifies optomechanical oscillations of hybrid systems \cite{sature2,sature3}, and can also influence macroscopic quantum properties of coupled resonators \cite{sature1}. In our concerned system of three coupled optical resonators, the time-dependent net optical gain rate can affect the intracavity fields. The indicator for the system to evolve to a desired dark mode is whether the field amplitude in the first cavity will finally tend to $|\langle\hat{a}_1\rangle|=0$, where $\langle\hat{a}_1\rangle=A_1e^{-i\omega_pt}$. Without a loss of the described features, we approximate with $\gamma_2=0$ in our model calculations because this loss can be compensated by a higher $\kappa_{2,0}$, so that the net gain rate keeps to be $\kappa_{2}(t)>0$. We let the system evolve under two conditions: (1) the resonant input for three identical resonance cavity frequencies; (2) $J_2=\sqrt{\kappa_{2,0}\kappa_{3}}$ for the chosen system parameters. For a constant $\kappa_2=\kappa_{2,0}$, there will emerge an exact $|\langle\hat{a}_1\rangle|=0$ under these conditions, as we illustrate in the main text, but this result cannot be guaranteed with a variable $\kappa_2(t)$ in Eq. (\ref{gain}). Through the numerical simulations based on Eq. (\ref{equations}), 
	we find that three factors affect such processes. The first one of the most obvious is the saturation intensity $I_S$. 
	Since the gain rate is a constant $\kappa_{2,0}$ in the limit $I_S\rightarrow\infty$, it is better to have an $I_S$ as higher as possible, as seen from Fig. S-1(a). Due to a lower $I_S$, the system will evolve to a state obviously deviating from $|\langle\hat{a}_1\rangle|=0$. However, this defect can be alleviated by the two other factors that are relevant to the dynamical evolution of the system. In Fig. S-1(b), we fix the gain saturation intensity $I_S$ together with other parameter, but the initial rate $\kappa_{2,0}$ can be varied. A smaller $\kappa_{2,0}$ is beneficial to realizing a state close to the dark mode with $|\langle\hat{a}_1\rangle|=0$. Increasing the inter-cavity coupling $J_1$ between cavity $1$ and cavity $2$ can also help the realization of a better dark mode; see the comparison in Fig. S-1(c) showing that a larger $J_1$ makes the system evolve to a more suppressed $|\langle\hat{a}_1\rangle|$. A better dark mode in approximation can be approached with the suitable parameters. For example, with this set of parameters, $\kappa_{2,0}=10^{-3}\,\mathrm{MHz}$, $I_S=10^8$, $J_1=10\,\mathrm{MHz}$, $\kappa_{1}=2\kappa_{ex}=2\,\mathrm{MHz}$, and $\kappa_3=1\,\mathrm{MHz}$, the system under the above-mentioned conditions and a resonantly input field with $\varepsilon_p=10^4\,\sqrt{\mathrm{MHz}}$ will have the finally evolved photon number $\langle \hat{a}_1^\dagger\hat{a}_1\rangle=7.69\times 10^{-6}$ in cavity $1$. 
	
	An interesting point is that the exact dark mode under an arbitrary gain saturation can be also realized by adjusting the coupling $J_2$. With the computation platforms such as Wolfram Mathematica, we can obtain the steady solution to Eq. (\ref{equations}) (after setting the left hand sides of the mean-field version of the equations to be zero) in a highly complex form of continued fraction. The coupling strength $J_2$ at the point of $|\langle\hat{a}_1\rangle|=0$ can be thus numerically found. For example, given a setup with $\kappa_{2,0}=0.01\,\mathrm{MHz}$, $I_S=10^8$, $J_1=10\,\mathrm{MHz}$, $\kappa_{3}=1\,\mathrm{MHz}$, and $\kappa_{1}=2\kappa_{ex}=2\,\mathrm{MHz}$, the coupling rate $J_2$ between cavity $2$ and $3$ is numerically found at $J_2=0.099\,\mathrm{MHz}$ to achieve the exact $|\langle\hat{a}_1\rangle|=0$ under a resonantly inputting field with $\varepsilon_p=10^4\,\sqrt{\mathrm{MHz}}$. This coupling strength is not far away from the value $J_2=\sqrt{\kappa_{2,0}\kappa_{3}}$.  
	
	The optical amplification noise is negligible in the realistic situations, though the similar noise demonstrates significance effect in some aspects of PT-symmetric quantum optics (such as quantum Fisher information \cite{noise-pt}). It is easy to see that after the system stabilizes, the gain noise term in Eq. (\ref{equations}) acts like an independent random drive aside from the coherent drive of the input, generating the extra photons. This amount of extra photons can be accurately calculated in a stabilized state described by the linear differential equations of the coupled modes. Even for a real-time process due to $\kappa_{2}(t)$, the evolving photon number due to the optical gain noise can be also calculated with a method used for simulating the dynamical processes of optomechanical cooling \cite{noise-formalism}. Here we provide an evaluation of the noise contribution with the example in Fig. 1(b) of the main text, in which the extra photons induced by the gain noise to cavity $1$ 
	is found to have the number
	$$ \langle \hat{a}_1^\dagger\hat{a}_1\rangle_n=\left|C_{1,2}\frac{g_{2,s}}{\lambda_1}\right|^2=\left|(0.707-0.186i)\frac{5.195\times 10^6}{1.997\times 10^7}\right|^2=3.6\times 10^{-2},$$
	where $C_{1,2}$ is the $(1,2)$-th element of the rotation matrix for diagonalizing the matrix $\textbf{M}$ defined in Eq. (3) of the main text (in a reference frame at the pump frequency), $g_{2,s}=\kappa_{2,s}+\gamma_2$ the gain rate after its saturation (the three resonators are assumed to have the same damping rate so that $\gamma_2=5$ MHz), and $\lambda_1$ the first eigenvalue of the matrix $\textbf{M}$. An alternative way for verifying the insignificance of the noise effect is to see on which power level the contribution of the pump drive term $\sqrt{2\kappa_{ex}}\varepsilon_{p}e^{-i\omega_pt}$ becomes comparable to that of the noise term $\sqrt{2g_{2}}\hat{\xi}^\dagger$, i.e. the pump power corresponding to $\varepsilon_{p}=1 \,\sqrt{\mathrm{Hz}}$ in Eq. (\ref{equations}). For the pump laser in the bandwidth of $\lambda=1550$ nm, this power is approximately $1.28\times 10^{-19}$ Watt, much lower than that of any feasible input. Experimentally the gain noise contribution is not observable even with a gain rate on the order of GHz \cite{ex-pt1}.          
	
\section{B. Induced Transparency with Different Structures}

	Without loss of generality, we here consider that the resonators, which are coupled one after another to form a string, have the same resonance frequency $\omega_{0}$ and the net gain rate $\kappa_2$ used in some of the structures is a constant. 
	The induced transparency phenomenon starts from two coupled passive resonators with their damping rates $\kappa_{1}$ and $\kappa_2$, respectively. From the corresponding dynamics, one finds the susceptibility of the structure as 
	\begin{eqnarray}
		\varepsilon_{T}=\frac{\sqrt{2\kappa_{ex}}A_{1}}{\varepsilon_{p}}=\dfrac{2\kappa_{ex}}{\kappa_{1}-ix+\dfrac{J_{1}^{2}}{\kappa_{2}-ix}},\quad 
	\end{eqnarray}
	where $x=\omega_{p}-\omega_{0}$ and the notation is consistent with our form of the dynamical equations in the main text. 
	The above form exactly corresponds to that of a three-level atomic medium for atomic EIT \cite{review}, and it leads to the transmission rate $T=C^2/(C+1)^2$ at the center of transparency window, where the cooperativity parameter is given as $C=J_1^2/(\kappa_{1}\kappa_2)$ (reformulated according to our form of dynamical equations). Its level of transparency continuously improves with the increased inter-cavity coupling $J_1$, analogous to the EIT of three-level atomic medium where the Rabi frequency of a much stronger coupling field plays the same role as that of $J_1$ here, but it can never reach a perfect intracavity field cancellation to $\varepsilon_{T}=0$, due to the fact $\kappa_{1},\kappa_{2}\neq 0$. The complete transparency ($100$\% transmission) with this structure can be only asymptotically approached with nearly lossless resonators or huge inter-cavity coupling strength. If the second resonator is replaced by an active one with a constant net gain rate $\kappa_2$, like in PT symmetric photonics \cite{pt-3,pt-7}, the susceptibility will become 
	\begin{eqnarray}
		\varepsilon_{T}=\dfrac{2\kappa_{ex}}{\kappa_{1}-ix+\dfrac{J_{1}^{2}}{-\kappa_{2}-ix}},\quad 
	\end{eqnarray}
	where the parameter $\kappa_{2}$ changes the sign. In this case, there only exists one trivial possibility of complete transparency at $x=0$, when the gain rate $g_2$ of the second resonator gets equal to its intrinsic damping rate $\gamma_2$ such that $\kappa_{2}=g_2-\gamma_2=0$. It should be noted that the dispersion properties reflected by the imaginary part of the susceptibility cannot be modified by the added gain, so that the induced transparency and field amplification work separately in this configuration.
	
	The next step is to couple one more resonator to the above-mentioned structures with the coupling strength $J_2$. If all of them
	are passive ones, we will have the following generalization,  
	\begin{eqnarray}
		\varepsilon_{T}=\dfrac{2\kappa_{ex}}{\kappa_{1}-ix+\dfrac{J_{1}^{2}}{\kappa_{2}-ix+\dfrac{J_{2}^{2}}{\kappa_{3}-
					ix}}}\quad 
	\end{eqnarray}
	from Eq. (S-4). This structure has the similar properties to those of two coupled passive resonators. However, if cavity $2$ is replaced by an active one such that $\kappa_2$ changes the sign to $-\kappa_2$, i.e.,
	\begin{eqnarray}
		\varepsilon_{T}=\dfrac{2\kappa_{ex}}{\kappa_{1}-ix+\dfrac{J_{1}^{2}}{-\kappa_{2}-ix+\dfrac{J_{2}^{2}}{\kappa_{3}-
					ix}}},\quad 
	\end{eqnarray}
	the part below the factor $J_1^2$ in the above expression can totally vanish under the conditions $J_2=\sqrt{\kappa_{2}\kappa_{3}}$ and $x=0$. It realizes a special dark mode with $\langle\hat{a}_1\rangle=0$, which fundamentally advances the induced transparency with coupled optical resonators. In contrast to the structures of coupled passive resonators, moreover, the induced transparency with the $100$\% transmission exists irrespective of the coupling strength $J_1$ to the first resonator.
	
	Successively one can add one more passive resonator to this string of three coupled resonators, with the dynamical matrix in Eq. (3) of the main text being 
	generalized to 
	\begin{equation}
		\textbf{M}=\left(
		\begin{array}
			[c]{cccc}%
			-\kappa_{1}-i\omega_{0} & iJ_{1} & 0 & 0\\
			iJ_{1} & \kappa_{2}-i\omega_{0} & iJ_{2} & 0\\
			0 & iJ_{2} & -\kappa_{3}-i\omega_{0} & iJ_{3}\\
			0 & 0 & iJ_{3} & -\kappa_{4}-i\omega_{0}
		\end{array}
		\right).
		\label{matrix}
	\end{equation}
	Then, the susceptibility will be further generalized to
	\begin{eqnarray}
		\varepsilon_{T}=\dfrac{2\kappa_{ex}}{\kappa_{1}-ix+\dfrac{J_{1}^{2}}{-\kappa_{2}-ix+\dfrac{J_{2}^{2}}{\kappa_{3}-
					ix+\dfrac{J_{3}^{2}}{\kappa_{4}-ix}}}}.\quad 
		\label{4 resonators}
	\end{eqnarray}
	A complete transparency can be thus realized after meeting the conditions:
	\begin{eqnarray}
		J_{2}&=&\sqrt{\dfrac{\kappa_{2}(J_{3}^{2}+\kappa_{3}\kappa_{4})}{\kappa_{4}}}\notag\\
		\omega_{p}&=&\omega_{0},
		\label{distribution-4}
	\end{eqnarray}
	such that the part below the factor $J_1^2$ in Eq. (\ref{4 resonators}) vanishes.
	Hence, the occurrence of a dark mode in cavity $1$ is completely determined by the parameters of the rest resonators---the coupling strengths $J_2$ and $J_3$ for the two pairs of the other resonators, together with the damping rates $\kappa_3$, $\kappa_4$ and gain rate $\kappa_2$ of them. Different from the possible field cancellation to $\langle\hat{a}_1\rangle=0$ due to two active-passive-coupled resonators, the net gain rate $\kappa_{2}$ in this structure can have arbitrary value, as long as it locates on the submanifold determined by Eq. (\ref{distribution-4}), which is possible by means of the adjustment of the other parameters. The generalization to even more resonators in a string is straightforward. Even if the structure is further extended beyond a string of coupled resonators, it will be still feasible to have the modified forms of continued fraction, similar to a practice in the atomic EIT involving multiple energy levels \cite{EIT-review}. These continued fractions give us the more complex submanifolds, on which the complete transparency through the structures can be realized while the parameter spaces grow to 
	the higher and higher dimensions.

	\section{C. Cavity Dark mode and Effective Hamiltonian}
	
	The effective Hamiltonian in Eq. (1) of the main text takes the form,
	\begin{eqnarray}
		\hat{H}_{eff}&=&(-\Delta_{1}-i\kappa_{1})\hat{a}_{1}^{\dagger}\hat{a}_{1}+(-\Delta_{2}+i\kappa_{2})\hat{a}_{2}^{\dagger}\hat{a}_{2}+(-\Delta_{3}-i\kappa_{3})\hat{a}_{3}^{\dagger}\hat{a}_{3}-J_{1}(\hat{a}_{1}^{\dagger}\hat{a}_{2}+\hat{a}_{2}^{\dagger}\hat{a}_{1})\notag\\
		&-&J_{2}(\hat{a}_{2}^{\dagger}\hat{a}_{3}+\hat{a}_{3}^{\dagger}\hat{a}_{2})+i\sqrt{2\kappa_{ex}}\varepsilon_{p}(\hat{a}_{1}^{\dagger}-\hat{a}_{1}),\quad   
	\end{eqnarray}%
	in the reference frame rotating at the input field frequency $\omega_p$.
	Like a laser beam successively split by a group of beamsplitters, the cavity field state due to the action of the above Hamiltonian is a product of coherent states $\left|\alpha_{1},\alpha_{2},\alpha_{3}\right\rangle=\left|A_1,A_2,A_3\right\rangle$, with $\hat{a}_{i}\left|A_{i}\right\rangle=A_{i}\left|A_{i}\right\rangle$ for $i=1,2,3$. Under the condition $J_2=\sqrt{\kappa_{2}\kappa_{3}}$, the evolved intracavity fields are the following: 
	\begin{eqnarray}
		A_{1}&=&0\notag\\
		A_{2}&=&\dfrac{i}{J_{1}}\times\sqrt{2\kappa_{ex}}\varepsilon_{p}\notag\\
		A_{3}&=&-\sqrt{\frac{\kappa_2}{\kappa_3}}\frac{1}{J_1}\times\sqrt{2\kappa_{ex}}\varepsilon_{p},
		\label{sol}
	\end{eqnarray}%
	for three cavities with the same resonance frequency $\omega_{1}=\omega_{2}=\omega_{3}=\omega_{0}$, which are under the driving field of the detuning $x=\omega_{p}-\omega_{0}=0$. The effective Hamiltonian will be thus reduced to
	\begin{eqnarray}
		\hat{H}_{eff}&=&-i\kappa_{1}\hat{a}_{1}^{\dagger}\hat{a}_{1}+i\kappa_{2}\hat{a}_{2}^{\dagger}\hat{a}_{2}-i\kappa_{3}\hat{a}_{3}^{\dagger}\hat{a}_{3}-J_{1}(\hat{a}_{1}^{\dagger}\hat{a}_{2}+\hat{a}_{2}^{\dagger}\hat{a}_{1})\notag\\
		&-&J_{2}(\hat{a}_{2}^{\dagger}\hat{a}_{3}+\hat{a}_{3}^{\dagger}\hat{a}_{2})+i\sqrt{2\kappa_{ex}}\varepsilon_{p}(\hat{a}_{1}^{\dagger}-\hat{a}_{1}).\quad 
		\label{H1}  
	\end{eqnarray}%
	
	When the system is in the state $\left|0,A_{2},A_{3}\right\rangle$, where $A_2$ and $A_3$ are given in Eq. (\ref{sol}), the action of the Hamiltonian in Eq. (\ref{H1}) will be
	\begin{eqnarray}
		&&\hat{H}_{eff}\left|0,A_{2},A_{3}\right\rangle=\left(i\kappa_{2}\hat{a}_{2}^{\dagger}\hat{a}_{2}-i\kappa_{3}\hat{a}_{3}^{\dagger}\hat{a}_{3}-J_{1}\hat{a}_{1}^{\dagger}\hat{a}_{2}-J_{2}(\hat{a}_{2}^{\dagger}\hat{a}_{3}+\hat{a}_{3}^{\dagger}\hat{a}_{2})+i\sqrt{2\kappa_{ex}}\hat{a}_{1}^{\dagger}\varepsilon_{p}\right)\left|0,A_{2},A_{3}\right\rangle,\nonumber\\
		&=&  	\underbrace{(-J_{1}\hat{a}_{1}^{\dagger}\hat{a}_{2}+i\sqrt{2\kappa_{ex}}\hat{a}_{1}^{\dagger}\varepsilon_{p})\left|0,A_{2}\right\rangle}_{=0}
		+\underbrace{\left(i\kappa_{2}\hat{a}_{2}^{\dagger}\hat{a}_{2}-i\kappa_{3}\hat{a}_{3}^{\dagger}\hat{a}_{3}-J_{2}(\hat{a}_{2}^{\dagger}\hat{a}_{3}+\hat{a}_{3}^{\dagger}\hat{a}_{2})\right)}_{\hat{H}_{R}}\left|A_{2}, A_{3}\right\rangle,
		\label{decompose}
	\end{eqnarray}%
	where we have used
	\begin{eqnarray}
		(-J_{1}\hat{a}_{1}^{\dagger}\hat{a}_{2}+i\sqrt{2\kappa_{ex}}\hat{a}_{1}^{\dagger}\varepsilon_{p})\left|0,A_{2}\right\rangle= (-J_{1}A_{2}+i\sqrt{2\kappa_{ex}}\varepsilon_{p})\left|1,A_{2}\right\rangle=0
	\end{eqnarray}%
	according to the solution in Eq. (\ref{sol}).
	
	The remaining Hamiltonian in Eq. (\ref{decompose}) can be written as
	\begin{eqnarray}
		\hat{H}_R&=&i\kappa_{2}\hat{a}_{2}^{\dagger}\hat{a}_{2}-i\kappa_{3}\hat{a}_{3}^{\dagger}\hat{a}_{3}-J_{2}(\hat{a}_{2}^{\dagger}\hat{a}_{3}+\hat{a}_{3}^{\dagger}\hat{a}_{2})=i\hat{b}^{\dagger}\hat{d}
	\end{eqnarray}%
	where
	\begin{eqnarray}
		\hat{b}=-i\sqrt{\kappa_{2}}\hat{a}_{2}-\sqrt{\kappa_{3}}\hat{a}_{3}\notag\\
		\hat{d}=-i\sqrt{\kappa_{2}}\hat{a}_{2}+\sqrt{\kappa_{3}}\hat{a}_{3}
	\end{eqnarray}%
	Acting the operator $\hat{d}$ on the state $\left|A_{2}, A_{3}\right\rangle$, we have
	\begin{eqnarray}
		\hat{d}\left|A_{2},A_{3}\right\rangle=(-i\sqrt{\kappa_{2}}A_{2}+\sqrt{\kappa_{3}}A_{3})\left|A_{2}, A_{3}\right\rangle=0,
	\end{eqnarray}%
	due to the solution in Eq. (\ref{sol}) again.
	Hence, the state $\left|0,A_{2},A_{3}\right\rangle$ is the one with the zero eigenvalue for the Hamiltonian $\hat{H}_{eff}$, 
	i.e., $\hat{H}_{eff}\left|0,A_{2},A_{3}\right\rangle=0$ for the $A_2$ and $A_3$ in Eq. (\ref{sol}). We call the operator $\hat{d}$ the dark-state operator and $\hat{b}$ the bright-state operator. They are not orthogonal, i.e., $[b,d^\dagger]\neq 0$, as a characteristic of non-Hermitian optical systems (see, e.g., Supplementary Materials of Ref. \cite{non-orthorgonal}).
	
	Confined to the state $\left|0,A_{2}\right\rangle$ of the first two cavities, only the two terms, $-J_{1}\hat{a}_{1}^{\dagger}\hat{a}_{2}$ and $i\sqrt{2\kappa_{ex}}\hat{a}_{1}^{\dagger}\varepsilon_{p}$, in Eq. (\ref{decompose}) can affect the photon number in cavity $1$. However, their joint action exactly cancel the contribution of each other according to the solution in Eq. (\ref{sol}), implying a perfect destructive interference of the field coupled back from cavity $2$ with the inputted field. Then, we define two Hamiltonians $\hat{H}_1=i\sqrt{2\kappa_{ex}}\varepsilon_{p}(\hat{a}_{1}^{\dagger}-\hat{a}_1)$ and  $\hat{H}_2=-J_1(A_2\hat{a}_1^\dagger+A_2^{\ast}\hat{a}_1)$ of the equivalent effects for these two sources of interference. 
	The joint action, $\langle 1|\hat{H}_1+\hat{H}_2|0\rangle=0$, forbids the excitation of cavity $1$, when the setup is under the conditions $J_2=\sqrt{\kappa_{2}\kappa_{3}}$ and $x=0$. This result is irrespective of the parameters $J_1$ and $\sqrt{2\kappa_{ex}}\varepsilon_{p}$ that are directly relevant to the field in cavity $1$. 
	
	\begin{figure}[t]
		\centering
		\includegraphics[width=0.9\textwidth]{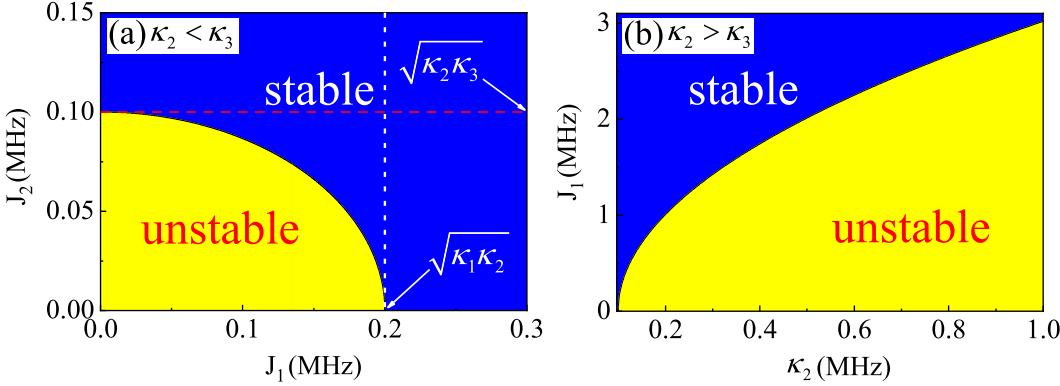}\quad 
		\caption{(a) One example of the dynamical regimes for the setups with $\kappa_2<\kappa_3$. The fixed system parameters 
			are chosen as $\kappa_{1}=2\kappa_{ex}=40\,\mathrm{MHz}$, $\kappa_{2}=0.001\,\mathrm{MHz}$, and $\kappa_{3}=10\,\mathrm{MHz}$. The two coupling strengths $J_1$ and $J_2$ are tunable, with their two particular values $\sqrt{\kappa_{1}\kappa_2}=0.2\,\mathrm{MHz}$ and $\sqrt{\kappa_{2}\kappa_3}=0.1\,\mathrm{MHz}$ for the determination of dynamical stability. (b) The preservation of dynamical stability with the enhanced coupling $J_1$, for a setup with $\kappa_2>\kappa_3$. Here the fixed parameters are set at $\kappa_{1}=2\kappa_{ex}=10\,\mathrm{MHz}$, $\kappa_{3}=0.1\,\mathrm{MHz}$, and the coupling $J_2$ is variable to be at the values $\sqrt{\kappa_{2}\kappa_3}$. The horizontal axis starts after $\kappa_{2}=0.1\,\mathrm{MHz}$, the value of $\kappa_{3}$.}
		\label{FigS2}%
	\end{figure}

	\subsection{D. Dynamical Stability and Experimental Feasibility}
	
	Modeled by Eq. (\ref{equations}), the structure driven by the input field stabilizes after the gain saturation, so that 
	Eq. (\ref{equations}) will finally reduce to a system of linear differential equations with the constant coefficients. 
	Depending on the chosen system parameters, some of the stabilized systems demonstrate the phenomenon of complete transparency, but the others cannot. Here we provide a categorization in the parameter space of the structure.
	
	The above-mentioned feasibility is relevant to the dynamical stability of the system with a constant net gain rate $\kappa_{2}$ ($I_S\rightarrow \infty$), and it is determined by the eigenvalues of matrix $\textbf{M}$ defined in Eq. (3) of the main text. The situations of $\kappa_{2}<\kappa_{3}$ have the totally different stable regimes from those of $\kappa_{2}>\kappa_{3}$. To the former, there are the following results for the resonant input fields: 
	$$ \text{If}~J_1 \geq \sqrt{\kappa_1\kappa_{2}},~ \text{the real parts of the eigenvalues of}~ \textbf{M}\leq 0~\text{for any}~J_2; $$
	$$ \text{If}~J_1 < \sqrt{\kappa_1\kappa_{2}},~ \text{the real parts of the eigenvalues of}~ \textbf{M}\leq 0~\text{when}~J_2>\sqrt{\frac{\kappa_1\kappa_2\kappa_3-J_1^2\kappa_3}{\kappa_1}}; $$
	$$ \text{If}~J_2 \geq \sqrt{\kappa_2\kappa_{3}},~ \text{the real parts of the eigenvalues of}~ \textbf{M}\leq 0~\text{for any}~J_1; $$
	$$ \text{If}~J_2 < \sqrt{\kappa_2\kappa_{3}},~ \text{the real parts of the eigenvalues of}~ \textbf{M}\leq 0~\text{when}~J_1>\sqrt{\frac{\kappa_1\kappa_2\kappa_3-J_2^2\kappa_1}{\kappa_3}}. $$
	These stable regimes distribute as those exemplified in Fig. \ref{FigS2}(a). Here we assume that only the two coupling strengths $J_1$ and $J_2$ are tunable by the relative distances between the resonators, with the practical examples for microspheres \cite{microsphere} and microtoroid resonators \cite{coupling} (the range of tuning $J_1$ and $J_2$ of this type of microresonators can be over $1$ GHz), while all other system parameters are fixed after the resonators are fabricated. Fig. 4 in the main text presents a process of tuning the coupling $J_2$ while the value of $J_1$ is fixed on the left side of the vertical line of $J_1=\sqrt{\kappa_1\kappa_2}$ in Fig. \ref{FigS2}(a). The coupling $J_2$ can be continuously tuned downwards across the horizontal line of $J_2=\sqrt{\kappa_2\kappa_3}$ in Fig. \ref{FigS2}(a), on which there is a complete cancellation of the field in cavity $1$, outputting the field from lower than the $100\%$ transmission to higher than the $100\%$ transmission. After the coupling $J_2$ touches the lower boundary, the system will enter the regime of dynamical instability, where the optical gain saturation will play a significant role so that the finally stabilized gain rate will be much lower than the original $\kappa_{2,0}$. On the vertical line of $J_1=\sqrt{\kappa_1\kappa_2}$, the output field intensity at the transparency window center will tend to infinitely high as the coupling $J_2$ is approaching to zero, but the associate transparency window will be vanishing at the same time. The output will always tend to infinity if the system approaches the boundary between the stable and unstable regime, except for the point $(J_1,J_2)=(0,\sqrt{\kappa_{2}\kappa_{3}})$. Going to the right side of the vertical line $J_1=\sqrt{\kappa_1\kappa_2}$, the system will behave like two coupled resonators in a PT-symmetric regime when one of the couplings is reduced to $J_2=0$, and the output with a fixed $J_1$ reaches the maximum there. It is clearly seen from Fig. S-2(a) that the smaller the net gain rate $\kappa_{2}$ is (with the fixed other built-in parameters), the smaller the unstable regime of the system will be shrunk to.
	
	If the built-in net gain rate $\kappa_{2}$ is higher than the damping rate $\kappa_3$ in a neighboring resonator, the system will have to reduce the regimes of stability. In this case, the system stability after an increased gain rate $\kappa_2$ should be preserved by a correspondingly enhanced inter-cavity coupling $J_1$; see Fig. \ref{FigS2}(b). For the example in Fig. \ref{FigS2}(b), the coupling $J_1$ must be increased to $3.017$ MHz when the net gain rate $\kappa_2$ reaches $1$ MHz (larger than the used $\kappa_3=0.1$ MHz). Because the situations of $\kappa_{2}>\kappa_{3}$ are less important, we will not elaborate on the distribution of associated dynamical regimes. 
	
	The system dynamics can be as well understood when there is an obvious gain saturation effect. For any possible gain saturation intensity $I_S$, the gain rate $g_2$ after saturation (even in the processes of dynamically unstable regimes) will become a lowered constant, so that the system can be nonetheless described by a group of linear differential equations for its coupled intracavity modes. The saturated gain rate, $g_{2,s}=\kappa_{2,0}/\left(1+\frac{<\hat{a}^\dagger_{2}\hat{a}_{2}>_s}{I_S}\right)$ ($<\hat{a}^\dagger_{2}\hat{a}_{2}>_s$ is the stabilized photon number in the active resonator), will be lowered from the original $\kappa_{2,0}$, and the complete transparency point will be moved to another coupling value $J_2=\sqrt{(g_{2,s}-\gamma_2)\kappa_3}$, as long as the saturated net gain rate, $g_{2,s}-\gamma_2$, is still positive. In the dynamically stable regimes determined by the original gain rate $\kappa_{2,0}$, i.e., the net gain rate $\kappa_{2}$ in Fig. \ref{FigS2} is replaced by another constant $\kappa_{2,0}-\gamma_{2}$, the intracavity field intensity or the photon number $<\hat{a}^\dagger_{2}\hat{a}_{2}>$ in the active cavity will cause a rather limited shift of the gain rate $g_2$ from the original $\kappa_{2,0}$, so that the complete transparency can be achieved by a relatively fine tuning of the coupling $J_2$ to a smaller value; see an example in Sec. I. It is why the complete transparency can be always reached at a coupling $J_2$ in such dynamically stable regimes, given any feasible gain saturation intensity $I_S$. 
	An interesting feature is that tuning the coupling $J_2$ around the point of complete transparency will not lead to a drastic change of the photon number $<\hat{a}^\dagger_{2}\hat{a}_{2}>$ in the active resonator. Especially, at the exact point of complete transparency, there is a counter-intuitive photon number $<\hat{a}^\dagger_{2}\hat{a}_{2}>_s=2\kappa_{ex}(\varepsilon_{p}/J_1)^2$ [from Eq. (\ref{sol})] that is independent of the coupling $J_2$ and the net gain rate, and it provides a mechanism to reduce the gain saturation effect by an enhanced coupling $J_1$ [exhibited by the results in Fig. \ref{FigS1}(c)]. On the other hand, the optical gain will drop tremendously due to a quick growth of the cavity photon number $<\hat{a}^\dagger_{2}\hat{a}_{2}>$ in 
	the dynamically unstable regime, and the stabilized gain rate $g_{2,s}$ will be thus far away from the original $\kappa_{2,0}$, making it impossible to reach the complete transparency point by tuning the coupling $J_2$.    
	
	In summary, the phenomenon of complete transparency through the concerned structure of three active-passive-coupled optical resonators should be observed in the dynamically stable regimes determined by its built-in gain rate $\kappa_{2,0}$, which can become smaller with a gain saturation. The system's dynamical stability can be guaranteed if it exists with this original gain rate $\kappa_{2,0}$, because the net gain rate will be only lowered after its saturation so that the system becomes more stable. With a negligible gain saturation effect, i.e., the field intensity in the active resonator is not high enough to modify the gain rate obviously, the complete transparency point will be located by adjusting the distance between cavity $2$ and $3$ so that the corresponding coupling goes across the horizontal line $J_2=\sqrt{\kappa_{2}\kappa_3}$ in Fig. \ref{FigS2}(a), and the operation can be performed with any nonzero coupling $J_1$ between cavity $1$ and cavity $2$. Even if there exists an obvious gain saturation that shifts the stabilized optical gain rate, the point of complete transparency can still be located by a further tuning of the coupling $J_2$ to a bit smaller value. Moreover, the gain saturation effect can be well suppressed by enhancing the coupling strength $J_1$.  
\end{widetext}


\begin{thebibliography}{99}   
	
	\bibitem{Harris1990prl} S. E. Harris, J. E. Field, and A. Imamoglu, Nonlinear optical processes using electromagnetically induced transparency, Phys. Rev. Lett. 64, 1107 (1990).
	
	\bibitem{Boller1991prl} K.-J. Boller, A. Imamoglu, and S. E. Harris,  Observation of electromagnetically induced transparency, Phys. Rev. Lett. 66, 2593 (1991).
	
	\bibitem{Fleischhauer2005RMP} M. Fleischhauer, A. Imamoglu, and J. P. Marangos, Electromagnetically induced transparency: Optics in coherent media, Rev. Mod. Phys. 77, 633 (2005).
	
	\bibitem{Ham1997} B. S. Ham, P. R. Hemmer, and M. S. Shahriar, Efficient electromagnetically induced transparency in a rare-earth doped crystal, Opt. Commun. 144, 227 (1997).
	
	\bibitem{Hau1999Nat} L. V. Hau, S. E. Harris, Z. Dutton, and C. H. Behroozi, Light speed reduction to 17 metres per second in an ultracold atomic gas, Nature 397, 594 (1999).
	
	\bibitem{Al2004} G. Alzetta, S. Gozzini, A. Lucchesini, S. Cartaleva and T. Karaulanov, C. Marinelli, and L. Moi, Complete electromagnetically induced transparency in sodium atoms excited by a multimode dye laser, Phys. Rev. A 69, 063815 (2004).
	
	\bibitem{novikova-2012} I. Novikova, R. L. Walsworth, and Y. Xiao, Electromagnetically induced transparency-based slow and stored light in warm atoms, Laser Photon. Rev. 6, 333 (2012).
	
	\bibitem{amplified} D, Wang, C. Liu, C. Xiao, J. Zhang, H. M. M. Alotaibi, B. C. Sanders, L.-G. Wang, and S. Zhu, Strong coherent light amplification with double electromagnetically induced transparency coherences, Sci. Rep. 7, 5796 (2017).
	
	\bibitem{EIT-review} R. Finkelstein, S. Bali, O. Firstenberg, and I. Novikova, A practical guide to electromagnetically induced transparency in atomic vapor, New J. Phys. 25, 035001 (2023).
	
	
	\bibitem{artificial-atom} A. A. Abdumalikov, Jr., O. Astafiev, A. M. Zagoskin, Yu. A. Pashkin, Y. Nakamura, and J. S. Tsai, 
	Electromagnetically Induced Transparency on a Single Artificial Atom, Phys. Rev. Lett. 104, 193601 (2010).
	
	\bibitem{cc-1} A. Yariv, Y. Xu, R. K. Lee, and A. Scherer, Coupled-resonator optical waveguide: a proposal and
	analysis, Opt. Lett. 24, 711 (1999).
	
	\bibitem{cc-2} D. D. Smith, H. Chang, K. A. Fuller, A. T. Rosenberger, and R. W. Boyd, Coupled-resonator-induced transparency, Phys. Rev. A 69, 063804 (2004).
	
	\bibitem{cc-3} L. Maleki, A. B. Matsko, A. A. Savchenkov, and V. S. Ilchenko, Tunable delay line with interacting whispering-gallery-mode resonators, Opt. Lett. 29, 6 (2004).
	
	
	\bibitem{cc-5} Q. Xu, S. Sandhu, M. L. Povinelli, J. Shakya, S. Fan, and M. Lipson, Experimental Realization of an On-Chip All-Optical Analogue to Electromagnetically Induced Transparency, Phys. Rev. Lett. 96, 123901 (2006).
	
	\bibitem{microsphere} K. Totsuka, N. Kobayashi, and M. Tomita, Slow Light in Coupled-Resonator-Induced Transparency, Phys. Rev. Lett. 98, 213904 (2007).
	
	
	\bibitem{cc-6} Y.-F. Xiao, L. He, J. Zhu, and L. Yang, Electromagnetically induced transparency-like effect in a single polydimethylsiloxane-coated silica microtoroid, Appl. Phys. Lett. 94, 231115 (2009).
	
	\bibitem{cc-7} X. Yang, M. Yu, D.-L. Kwong, and C. W. Wong, All-Optical Analog to Electromagnetically Induced Transparency
	in Multiple Coupled Photonic Crystal Cavities, Phys. Rev. Lett. 102, 173902 (2009).
	
	\bibitem{cc-8} B. Peng, S. K. \"Ozdemir, W. Chen, F. Nori, and L. Yang, What is and what is not electromagnetically
	induced transparency in whispering-gallery microcavities, Nat. Commun. 5, 5082 (2014).
	
	\bibitem{cc-9} C. Wang, X. Jiang, W. R. Sweeney, C. W. Hsu, Y. Liu, G. Zhao, B. Peng, M. Zhang, L. Jiang, A. D. Stone, and L. Yang,
	Induced transparency by interference or polarization, Proc. Natl. Acad. Sci. 118, e2012982118 (2021).
	
	\bibitem{cc-10} D. Sugio, T. Manabe, K. Nakamura, T. Matsumoto, and M. Tomita, Observation of the transition from inverted coupled-resonator-induced transparency to inverted Autler-Townes splitting, Phys. Rev. A 107, 013110 (2023).
	
	\bibitem{report} R. W. Boyd and D. J. Gauthier, Photonics: transparency on an optical chip, Nature 441, 701 (2006).
	
	\bibitem{review} Y.-C. Liu, B.-B. Li, and Y.-F. Xiao, Electromagnetically induced transparency
	in optical microcavities, Nanophotonics 6, 789-811 (2017).
	
	\bibitem{trans-review} H. Qin, M. Ding, and Y. Yin, Induced Transparency with Optical Cavities, Adv. Photonics Research 1, 2000009 (2020).
	
	\bibitem{q-001} S. Zhang, D. A. Genov, Y. Wang, M. Liu, and X. Zhang, Plasmon-Induced Transparency in Metamaterials, Phys. Rev. Lett. 101, 047401 (2008).
	
	\bibitem{dipole} R. Puthumpally-Joseph, M. Sukharev, O. Atabek, and E. Charron, Dipole-Induced Electromagnetic Transparency,
	Phys. Rev. Lett. 113, 163603 (2014).
	
	\bibitem{q-0} C.-H. Dong, Z. Shen, C.-L. Zou, Y.-L. Zhang, W. Fu, and G.-C. Guo, Brillouin-scattering-induced transparency and non-reciprocal light storage, Nat. Commun. 6, 6193 (2015).
	
	\bibitem{q-01} J. Kim, M. C. Kuzyk, K. Han, H. Wang, and G. Bahl, Non-reciprocal Brillouin scattering induced transparency,
	Nature Physics 11, 275 (2015).
	
	\bibitem{q-1} Y. Zheng, J. Yang, Z. Shen, J. Cao, X. Chen, X. Liang, and W. Wan, Optically induced transparency in a micro-cavity, Light Sci. Appl. 5, 16072 (2016).
	
	\bibitem{q-2} J. Ma, J. Qin, G. T. Campbell, R. Lecamwasam, K. Sripathy, J. Hope, B. C. Buchler, and P. K. Lam, Photothermally induced transparency, Sci. Adv. 6, eaax8256 (2020).
	
	\bibitem {Huang2010_041803} G. S. Agarwal and S. Huang, Electromagnetically induced transparency in mechanical effects of light, Phys. Rev. A 81, 041803(R) (2010).
	
	\bibitem {Weis2010} S. Weis, R. Rivi$\grave{e}$re, S. Del$\acute{e}$glise, E. Gavartin, O. Arcizet, A. Schliesser, and T. J. Kippenberg, Optomechanically Induced Transparency, Science 330, 1520--1523 (2010).
	
	
	\bibitem {Safavi-Naeini2011} A. H. Safavi-Naeini, T. P. Mayer Alegre, J. Chan, M. Eichenfield, M. Winger, Q. Lin, J. T. Hill, D. E. Chang, and O. Painter, Electromagnetically induced transparency and slow light with optomechanics, Nature 472, 
	69 (2011).
	
	\bibitem{Huang2011} S. Huang and G. S. Agarwal, Electromagnetically induced transparency from two-phonon processes in quadratically coupled membranes, Phys. Rev. A 83, 023823 (2011).
	
	\bibitem{Chang2011njp} D. E. Chang, A. H. Safavi-Naeini, M. Hafezi, and O. Painter, Slowing and stopping light using an optomechanical crystal array, New J. Phys. 13, 023003 (2011).
	
	
	\bibitem{Xiong2012} H. Xiong, L.-G. Si, A.-S. Zheng, X. Yang, and Y. Wu, Higher-order sidebands in optomechanically induced transparency, Phys. Rev. A 86, 013815 (2012).
	
	\bibitem{Karuza2013} M. Karuza, C. Biancofiore, M. Bawaj, C. Molinelli, M. Galassi, R. Natali, P. Tombesi, G. Di Giuseppe, and D. Vitali, Optomechanically induced transparency in a membrane-in-the-middle setup at room temperature, Phys. Rev. A 88, 013804 (2013).
	
	
	\bibitem {Kronwald2013} A. Kronwald and F. Marquardt, Optomechanically Induced Transparency in the Nonlinear Quantum Regime, Phys. Rev. Lett. 111, 133601 (2013).
	
	
	\bibitem{Tarhan2013} D. Tarhan, S. Huang, and \"O. E. M\"ustecaplioğlu, Superluminal and ultraslow light propagation in optomechanical systems, Phys. Rev. A 87, 013824 (2013).
	
	
	\bibitem{Jing2015} H. Jing, S. K. \"Ozdemir, Z. Geng, J. Zhang, X.-Y. L\"u, B. Peng, L. Yang, and F. Nori, Optomechanically-induced transparency in parity-time-symmetric microresonators, Sci. Rep. 5, 9663 (2015).
	
	
	\bibitem{Lu2018} H. L\"u, C. Wang, L. Yang, and H. Jing, Optomechanically Induced Transparency at Exceptional Points, Phys. Rev. Applied 10, 014006 (2018).
	
	
	\bibitem{Yan2020pra} X. B. Yan, Optomechanically induced transparency and gain, Phys. Rev. A 101, 043820 (2020). 
	
	
	\bibitem{Yan2021PhysicaE} X. B. Yan, Optomechanically induced ultraslow and ultrafast light, Phys. E 131, 114759 (2021).
	
	\bibitem{cc-12} G. Strangi, A. De Luca, S. Ravaine, M. Ferrie, and R. Bartolino, Gain induced optical transparency in metamaterials, Appl. Phys. Lett. 98, 251912 (2011).
	
	\bibitem{mc1} X. Li, K. Qi, Y. Wu, X. Wu, M. Hu, Z. Li, Y. He, S. Ding, Z. Xie, H. Zhou, B. He, M. Xiao, and X. Jiang, 
	Generation of 2/3-octave-spanning visible Kerr soliton microcomb, Adv. Photon. 7, 056002 (2025).
	
	\bibitem{mc2} Q. Shi, J. Tian, S. Ding, Y. Wang, S. Lei, M. Zhang, W. Wan, X. Ji, B. He, M. Xiao, and X. Jiang, On-chip ultra-high-Q optical microresonators approaching the material absorption limit, Photonics Res. 13, 2409 (2025).
	
	\bibitem{mc3} X. Li, Y. Xie, S. Ding, M. Zhang, Y. He, B. He, M. Xiao, and X. Jiang, Observation of Kerr soliton microcomb locked with a phonon laser, Sci. Adv. 12, eaeb3400 (2026).
	
	\bibitem{pt-2} A. A. Zyablovsky, A. P. Vinogradov, A. A. Pukhov, A. V. Dorofeenko, and A. A. Lisyansky, PT-symmetry in optics, Phys.-Usp. 57, 1063 (2014).
	
	\bibitem{pt-3} V. V. Konotop, J. Yang, and D. A. Zezyulin, Nonlinear waves in PT-symmetric systems, Rev. Mod. Phys. 88, 035002 (2016).
	
	\bibitem{pt-4} J. Wen, X. Jiang, L. Jiang, and M. Xiao, Parity-time symmetry in optical microcavity systems, J. Phys. B: At. Mol. Opt. Phys. 51, 222001 (2018).
	
	\bibitem{pt-5} Ş. K. \"{O}zdemir, S. Rotter, F. Nori, and L. Yang, Parity–time symmetry and exceptional points in photonics, Nat. Mater. 18, 783 (2019). 
	
	\bibitem{pt-6} C. Wang, Z. Fu, W. Mao, J. Qie, A. D. Stone, and L. Yang, Non-Hermitian optics and photonics: from classical to quantum, Adv. Opt. Photon. 15, 442-523 (2023).
	
	\bibitem{pt-7} C. Zhang, M. Kim, Y.-H. Zhang, Y.-P. Wang, D. Trivedi, A. Krasnok, J. Wang, D. Isleifson, R. Roshko, and C.-M. Hu, 
	Gain-Loss Coupled Systems, APL Quantum 2, 011501 (2025).
	
	\bibitem{He2018prl} B. He, L. Yang, X. Jiang, and M. Xiao, Transmission Nonreciprocity in a Mutually Coupled Circulating Structure, Phys. Rev. Lett. 120, 203904 (2018).
	
	
	\bibitem{sature1} S. Vashahri-Ghamsari, B. He, and M. Xiao, Effects of gain saturation on the quantum properties of light in a non-Hermitian gain-loss coupler, Phys. Rev. A 99, 023819 (2019).
	
	
	\bibitem{sature2} Y. Xie, Z. Cao, B. He, and Q. Lin, PT-symmetric phonon laser under gain saturation effect, Opt. Express 28, 22580-22593 (2020).
	
	\bibitem{sature-nonreciprocal} Z. Cao, Y. Xie, B. He, and Q. Lin, Ultra-high optical nonreciprocity with a coupled triple-resonator structure, New J. Phys. 23 023010 (2021).
	
	\bibitem{sature3} X. Z. Hao, X. Y. Zhang, Y. H. Zhou, W. Li, S. C. Hou, and X. X. Yi, Gain-saturation-induced self-sustained oscillations in non-Hermitian optomechanics, Phys. Rev. A 103, 053508 (2021).
	
	
	\bibitem{gain-noise} G. S. Agarwal and K. Qu, Spontaneous generation of photons in transmission of quantum fields in PT-symmetric optical systems, Phys. Rev. A 85, 031802(R) (2012).
	
	\bibitem{sature4} S. Vashahri-Ghamsari and B. He, Gain Saturation Modified Quantum Noise Effect on Preparing a Continuous-Variable Entanglement, Photonics 9, 620 (2022).
	
	\bibitem{ex-pt1} L. Chang, X. Jiang, S. Hua, C. Yang, J. Wen, L. Jiang, G. Li, G. Wang, and M. Xiao, Parity–time symmetry and variable optical isolation in active–passive-coupled microresonators, Nat. Photon. 8, 524 (2014).
	
	
	
	\bibitem{ds0} J. Huang, C. Liu, X.-W. Xu, and J.-Q. Liao, Dark-mode theorems for quantum networks, arXiv:2312.06274.
	
	\bibitem{a-dm01} D. H. White, S. Kato, N. N\'emet, S. Parkins, and T. Aoki,
	Cavity dark mode of distant coupled atom-cavity systems,
	Phys. Rev. Lett. 122, 253603 (2019).
	
	\bibitem{a-dm02} S. Kato, N. N\'{e}met, K. Senga, S. Mizukami, X. Huang,
	S. Parkins, and T. Aoki, Observation of dressed states
	of distant atoms with delocalized photons in coupledcavities
	quantum electrodynamics, Nat. Commun. 10,
	1160 (2019).
	
	\bibitem{a-dm03} X. Zhang, Z. Yu, H. Zhang, D. Xiang, and H. Zhang, Cavity dark mode mediated by atom array without atomic scattering loss, Phys. Rev. Research 6, L042026 (2024).
	
	\bibitem{om-d1} C. Dong, V. Fiore, M. C. Kuzyk, and H. Wang, Optomechanical
	dark mode, Science 338, 1609 (2012).
	
	\bibitem{om-d2} J. T. Hill, A. H. Safavi-Naeini, J. Chan, and O. Painter,
	Coherent optical wavelength conversion via cavity optomechanics,
	Nat. Commun. 3, 1196 (2012).
	
	\bibitem{coupling} C. Yang, X. Jiang, Q. Hua, S. Hua, Y. Chen, J. Ma, and M. Xiao, Realization of controllable photonic molecule based on three ultrahigh‐Q microtoroid cavities, Laser Photon. Rev. 11, 1600178 (2017).
	
	\bibitem{nuc-trans} H.-b. Zhang, Y. Tang, and Y.-C. Liu, Nuclear spin induced transparency, Phys. Rev. Research 7, L012069 (2025).
	
	\bibitem{ex-pt2} B. Peng, Ş. K. \"Ozdemir, F. Lei, F. Monifi, M. Gianfreda, G. L. Long, S. Fan, F. Nori, C. M. Bender, and L. Yang, Parity–time-symmetric whispering-gallery microcavities, Nat. Phys. 10, 394 (2014).
	
	\bibitem{ep2} M. A. Miri and A. Al\'u, Exceptional points in optics and photonics, Science 363, eaar7709 (2019).
	
	\bibitem{ep3} J. Wiersig, Review of exceptional point-based sensors, Photonics Res. 8, 1457-1467 (2020).
	
	\bibitem{ep4} H. Meng, Y. S. Ang, and C. H. Lee, Exceptional points in non-Hermitian systems: Applications and recent developments,
	Appl. Phys. Lett. 124, 060502 (2024).
	
	\bibitem{book} C. W. Gardiner and P. Zoller, Quantum Noise, Springer-Verlag, Berlin Heidelberg, 2000.
	
	\bibitem{noise-formalism} B. He, L. Yang, Q. Lin, and M. Xiao, Radiation Pressure Cooling as a Quantum Dynamical Process, Phys. Rev. Lett. 118, 233604 (2017).
	
	\bibitem{noise-pt} W. Wang, Y. Zhai, D. Liu, X. Jiang, S. Vashahri-Ghamsari, and J. Wen, Quadrature-PT symmetry: Classical-to-quantum transition in noise fluctuations, Quantum Sci. Technol. 10, 025016 (2025).
	
	\bibitem{non-orthorgonal} B. He, L. Yang, and M. Xiao, Dynamical phonon laser in coupled active-passive microresonators, Phys. Rev. A 94, 031802(R) (2016).
	
\end{thebibliography}
\end{document}